\documentstyle[12pt]{article}
\begin{document}

\def \d {{\rm d}}

\title{Chaos in a modified H\'enon-Heiles system
describing geodesics in gravitational waves}

\author{
K. Vesel\'y,
\ J. Podolsk\'y\thanks{E--mail: {\tt podolsky@mbox.troja.mff.cuni.cz}}
\\ \\ Institute of Theoretical Physics,
 \\ Charles University,\\
V Hole\v{s}ovi\v{c}k\'ach 2, 18000 Prague 8, Czech Republic.\\ }

\maketitle

\baselineskip=21pt

\begin{abstract}
A Hamiltonian system with a modified H\'enon-Heiles potential
is investigated. This describes the motion of free test particles in
vacuum gravitational {\it pp}-wave spacetimes with both quadratic
(`homogeneous')
and cubic (`non-homogeneous') terms in the structural function.
It is shown that, for energies above a certain value, the motion is
chaotic in the sense that the boundaries separating the basins of possible
escapes
become fractal. Similarities and differences with the standard
H\'enon-Heiles and the monkey saddle systems are discussed.
The box-counting dimension of the basin boundaries is also
calculated.
\end{abstract}

\vfil\noindent

\vfil\noindent


\vfil
\eject

\section{Introduction}

Various open Hamiltonian systems in which unbounded orbits
exhibit a chaotic behaviour have been studied (in particular in
the context of chaotic scattering), see e.g. \cite{Rod}-\cite{OT}
and elsewhere. A standard method of investigation and
description of these systems is to examine a typical set of initial
conditions and observe which of the outcome possibilities is
assigned to each initial data. The presence of {\it fractal basin
boundaries} separating the possible escapes in the space of
initial conditions is a suitable and invariant characterization
of chaos \cite{Ott}-\cite{DFC}.

Among the simplest Hamiltonians of this type are those with two degrees
of freedom and polynomial potentials, in particular the
famous H\'enon-Heiles (HH) system \cite{HH} given by
\begin{equation}
H = \hbox{${1\over 2}$}(p_x^2+p_y^2)+V(x,y)\ ,
\label{Ham}
\end{equation}
where
\begin{equation}
V=V_{HH}(x,y) \equiv \textstyle{1\over 2}(x^2+y^2)+
\hbox{${1\over 3}$}(x^3-3xy^2)\ .
\label{VHH}
\end{equation}
In the literature, however, this is usually studied at energies below the
critical escape value $E_c={1\over 6}$, for which motions are
bounded. These are regular for lowest $E$ but become chaotic at
higher energies still obeying $E<E_c$. Chaos is also present
above $E_c$, where unbounded orbits choose between three possible
distinct escapes to infinity (see, e.g. \cite{ChRod}). The corresponding
fractal basin boundaries have recently been investigated in \cite{ML}.

In this paper a system with a potential similar to the standard
HH model (\ref{VHH}) is presented and studied. This naturally
arises in the context of test particle motion in plane-fronted
gravitational waves with parallel rays ({\it pp-}waves) \cite{KSMH}.
It is well-known that in the simplest (so called homogeneous)
{\it pp-}waves the geodesic motion is regular but, as shown recently
\cite{PoVe1}-\cite{PoVe3}, it is chaotic in
the non-homogeneous case.
It is the purpose of this paper to discuss in detail a combined
effect of both homogeneous and non-homogeneous gravitational-wave
contributions. In section 2 the geodesic equations of motion in
{\it pp-}wave spacetimes are presented in a suitable form in which the relation
to Hamiltonian dynamics is established. Specific energy manifolds are then
described and classified in section 3. The onset of chaos
investigated by numerical simulation of orbits in an interesting
medium energy region is described in section 4. Finally, in section~5
the box-counting dimension of the fractal basin boundaries is
calculated.

\section{Geodesics in gravitational {\it pp}-waves with both
homogeneous and non-homogeneous components}

The line element of the well-known class of gravitational {\it pp}-wave
solutions \cite{KSMH} in vacuum can be written
\begin{equation}
\d s^2=2\,\d\zeta \d\bar\zeta-2\,\d u\d v-(f+\bar f)\,\d u^2\ ,
\label{pp}
\end{equation}
where $f(u,\zeta)$ is an arbitrary complex function of the retarded time $u$
and the complex spatial coordinate $\zeta$. As shown in \cite{PoVe1},
\cite{PoVe2}, to determine the geodesic motion in these
space-times it is crucial to solve the complex equation
\begin{equation}
\ddot\zeta + (U^2/2)\,{\partial\bar f\over\partial\bar\zeta}=0\ ,
\label{E3}
\end{equation}
for $\zeta(\tau)$, where a dot denotes the derivative with respect to the
normalized `time' parameter $\tau$, and $U\equiv\dot u$ is a constant.
(The remaining functions
$u(\tau)$ and $v(\tau)$ can be found by subsequent integrations.)
Introducing two real coordinates by $x\equiv{\cal R}e\ \zeta$,
$y\equiv{\cal I} m\ \zeta$, the Eq. (\ref{E3}) can be considered
as a consequence of the Hamiltonian (\ref{Ham}) with the potential
$V(x,y,u)=(U^2/2)\,{\cal R}e f$.

The space-time (\ref{pp}) with the function $f$ linear in
$\zeta$ represents the trivial flat Minkowski universe. The quadratic
function $f\sim\zeta^2$ describes homogeneous {\it pp}-waves
(see \cite{KSMH} and references therein), in which the particle
motions are integrable. Geodesics in non-homogeneous {\it pp}-waves
given by $f\sim\zeta^n, n=3, 4, 5, \cdots$, have only recently been
investigated  \cite{PoVe1}, \cite{PoVe2}. The corresponding
polynomial potential $V(x,y)\sim{\cal R} e\, \zeta^n$ is called
an `$n$-saddle' (a `monkey saddle' for $n=3$). It has been found that
motion in these gravitational waves with constant `amplitude' are
unbounded with $n$ possible escapes to $|\zeta|\to \infty$, and
that the basin boundaries separating these outcomes are fractal.  For
non-constant amplitudes  describing sandwich and impulsive
gravitational waves a smearing of chaos is observed \cite{PoVe3}.

Now, it is natural and also physically interesting to investigate
the combined effect of quadratic
and higher polynomial terms in the function $f$. This is the main
purpose of our work here. The simplest situation of this type
occurs when  $f$ has both quadratic and cubic terms,
\begin{equation}
f(\zeta)=\alpha\zeta^2+{\textstyle{2\over3}}\beta\zeta^3\ ,
\label{f23}
\end{equation}
where $\alpha$ and $\beta$ are real positive constants.
With this, (\ref{E3}) reduces to
$\ddot\zeta + (\alpha\bar\zeta + \beta\bar\zeta^2)\,U^2 =0$, in which
the (non-trivial) constants $\alpha$, $\beta$ and $U$ can be removed
by a simple transformation
\begin{equation}
\zeta(\tau)\to{\beta \over \alpha}\zeta\left({\tau\over
U\sqrt\alpha}\right)\ .
\label{scaling}
\end{equation}
In the real coordinates $x$ and $y$, the equations of motion
then
take the form
\begin{equation}
\ddot x + x +x^2-y^2 =0\ , \qquad
\ddot y - y -2xy =0\ .
\label{E5}
\end{equation}
These equations can be considered to follow from
the Hamiltonian (\ref{Ham}) with the potential
\begin{equation}
V_m(x,y) = {\textstyle{1\over 2}}(x^2-y^2)+
{\textstyle{1\over 3}}(x^3-3xy^2)\ ,
\label{E6}
\end{equation}
which may be called a `modified H\'enon-Heiles potential' since
this has almost the form of the standard H\'enon-Heiles
potential (\ref{VHH}) with  the exception of the negative sign of the $y^2$
term. However, as will be shown below, this `small' difference
results in a behaviour which is qualitatively different from that observed
in the well-known HH system. Note that, from a previous
investigation of a generalized HH system
$V_g(x,y) = {1\over 2}(Ax^2+By^2)+{1\over 3}(Cx^3-3Dxy^2)$
`with adjustable coefficients', it is known that the system (\ref{E6})
(corresponding to the choice $A=C=D=1$, $B=-1$ in $V_g$) is {\it
not integrable} \cite{Tab}.

\section{Classification of possible orbits according to a
conserved energy}

First, we shall investigate specific properties of the polynomial
potential (\ref{E6}) and compare them with those of the standard
HH potential (\ref{VHH}) and the monkey saddle potential
$V_{ms}(x,y)={1\over 3}(x^3-3xy^2)$, in which the quadratic terms
are missing. There are two equilibrium saddle points of the
potential $V_m$ at $(x,y)=(0,0)$ and $(-1,0)$. This is different
from the HH potential (\ref{VHH}) for which there is an
elliptic point
at the origin (in addition to three saddle points), and from the
monkey saddle potential which has only one unstable equilibrium
point (at the origin).
The equipotentials $V_m(x,y)=E=\hbox{const.}$ of the potential (\ref{E6}) are
drawn in Fig.~1. Let us now briefly comment on various possible
situations shown in the figure:
\begin{itemize}
\item For $E<0$ the energy manifold $H=E$ consists of {\it three}
disjoint components (separated by a `forbidden' grey region, see Fig.~1),
each of which contains a single exit to infinity. The similar
behaviour is observed for negative energies also in the monkey saddle potential
$V_{ms}$ and in the HH system.
\item For $E=0$ the two components of the energy manifold join at
$(0,0)$, the third one remains separated.
\item For $0<E<{1\over 6}$ the energy manifold consists of {\it two}
disjoint components, one containing a single exit to infinity
($x\to-\infty$), the second containing two distinct exits
($y\to+\infty,\> y\to-\infty$). This situation does not occur in the
monkey saddle potential, in which all energy manifolds with $E>0$ are
connected and contain three exits to infinity. The energy manifolds
$0<E<{1\over 6}$ for the HH system consist of four disjoint components.
One corresponds to bounded orbits and the remaining three are unbounded,
each containing a single exit to infinity.
\item For $E={1\over 6}$ the two components of the energy manifold join
at the point $(-1,0)$.
\item For $E>{1\over 6}$ the energy manifold is connected and
contains three exits to infinity  in this case. The topology of energy
manifolds is the same for both the HH and the
monkey saddle systems.
\end {itemize}

The orbits in the negative energy region cannot choose between
different exits, so that no basin boundaries are present. On the
other hand, the high energy region ($E>{1\over 6}$) provides three
possible exits to infinity and numerical
simulations show that the basin boundaries are fractal indicating
the presence of chaos. These simulations give results very
similar to those for the monkey saddle presented in \cite{PoVe2}, or
for unbounded motions in the HH model \cite{ML}. This is not surprising
as the equipotentials of both the potential (\ref{E6}) and the standard
HH potential (\ref{VHH}) are well approximated by the monkey saddle
equipotentials  $V_{ms}=E$ for sufficiently high energies.

Thus, the most interesting energy region to be considered here is given
by the
condition $0<E<{1\over 6}$. Since the motion for $E<0$
is not chaotic while it is chaotic in the high energy region,
chaos has to appear somewhere between $E=0$ and $E={1\over 6}$.
This is similar to the standard HH model, where the low energy
orbits above $E=0$ are regular while they become chaotic for
higher energies still below $E={1\over 6}$. Such behaviour is not
present in the `pure cubic' monkey saddle potential $V_{ms}$ in
which one can clearly distinguish only two energy regions, negative and
positive. The existence of the medium energy region is caused by the
presence of
the quadratic terms in the potentials (\ref{VHH}) and (\ref{E6}).
In the following section we shall concentrate on the behaviour of
orbits in this interval.

\section{The onset of chaos in the medium energy region}

For an investigation of orbits in the medium energy region
$0<E<{1\over6}$ by a fractal method (which is useful for the characterization
of
a sensitive dependence on initial conditions in systems with unbounded
orbits) it is necessary to choose a suitable typical set of initial
data. Here we may concentrate
on motion in the  region with two
possible exits to infinity, $y\to\pm\infty$ (the right white region
in Fig.~1 d, e, f). Let us consider orbits starting at $\tau=0$ from rest
from the `right branch' of the level curve $B_r$ (the branch of
$V_m(x,y)=E$ with $x>0$), see Fig.~2 \  for typical trajectories
of this type. This choice of initial conditions is also useful for localizing
periodic orbits, which play an essential role in establishing the chaotic
behaviour in both the monkey-saddle system and the standard HH system
above the critical energy \cite{Rod}, \cite{ChRod}. We can conveniently
parametrize the starting points on $B_r$ using the polar coordinate
angle $\phi$ with respect to the origin (such that $|\phi|<{\pi\over 6}$).
Each orbit is then uniquely characterized by its energy $E$ and
a starting angle $\phi$.

In fact, we can concentrate on a smaller range of $E$ in the medium energy
region. It is easy to prove that for energies $0<E<{1\over 12}$ the
orbits starting from $B_r$ cannot cross the $x$-axis. Indeed, the
second equation of (\ref{E5}) can be rewritten as $\ddot y=y(1+2x)$.
Since $x>-{1\over 2}$ in the considered component of the energy manifold
for $E<{1\over 12}$, the acceleration $\ddot y$ has the same sign
as $y$. The basin boundary between the two possible exits (`up' and
`down') will therefore be given by $\phi=0$, as the orbits starting from
$B_r$ with $\phi>0$ will always escape directly to $y \to\infty$, while
the orbits with $\phi<0$ will always escape to $y \to-\infty$.
They never cross the $x$-axis, so the basin boundary
is {\it not} fractal. The unique orbit with $\phi=0$ is an unstable periodic
orbit $\Pi_0$ with trajectory along the $x$-axis.

Consequently, chaos has to appear in the energy region
${1\over12}<E<{1\over6}$. Numerical simulations show that
behaviour of orbits with small $\phi$ becomes much more complicated
for energies $E$ above a certain critical value $E_0\in[{1\over
12}, {1\over 6}]$. The value of $E_0$ will be discussed and specified
later. Typical orbits for $E=0.16$ are drawn in Fig. 2. In addition to
the unstable periodic orbit $\Pi_0$ along the $x$-axis,
there is an unstable periodic orbit $\Pi_3$ with $\phi>0$
(see the bold curves in Fig. 2). Another one $\Pi_2$ with $\phi<0$
(which is not shown) can be obtained by symmetry.

At this point it is convenient to introduce a simple notation.
As mentioned above, there are two possible escapes to infinity
for orbits in the region being discussed.
Let us assign them a symbol $j$, where $j=2$
for the exit with $y\to-\infty$, and $j=3$ for the exit with
$y\to\infty$ ($j=1$ is reserved for the exit $x\to-\infty$, which
is present for $E>{1\over 6}$, to
keep the notation consistent with \cite{PoVe2} and \cite{PoVe3}).
Also, let us denote by $\tau_s$ the time $\tau>0$ at which the orbit
reaches infinity (we put $\tau_s=\infty$ for orbits that never
escape to infinity).

We shall now investigate the detailed structure of the orbits for the
typical
energy $E=0.16$.  In Fig.~3 the functions $j(\phi)$
and $\tau_s(\phi)$ are plotted. (These are analogues of the scattering
function and the time delay, respectively.) The basin boundaries
separating possible outcomes in the initial condition state
space are clearly {\it fractal}, which is confirmed here by zooming in
up to the fourth level. This is an invariant evidence that the
behaviour of orbits for energy $E=0.16$ is chaotic. At each level of the
fractal in Fig.~3, there are three peaks of $\tau_s(\phi)$. The left
(middle, right) peak corresponds to orbits asymptotic (up to the
given level) to the periodic orbit $\Pi_2$ ($\Pi_0$, $\Pi_3$).
Observe that a fractal structure occurs in a region in which there are
{\it only two} possible outcomes.

Further observations analogous to those described in \cite{PoVe2},
\cite{PoVe3} could be made. Instead, we shall concentrate here on
the feature which is not present in the monkey saddle system,
namely the dependence of $j(\phi)$ on the energy $E$. In addition to
a single-parameter $\phi$-dependence in Fig.~3, it is necessary to
investigate the basins for the `up' ($j=3$) and `down' ($j=2$)
exits in the two-parameter region $\{ (E,\phi)\ |\ {1\over
12}<E<{1\over 6},|\phi|<\phi_{max}={\pi\over 6}\} $. The result
for $|\phi|<0.1$ is presented in Fig.~4, in which black color
denotes $j=3$ and white corresponds to $j=2$. As expected,
the only basin boundary for $E$ below the
critical energy $E_0\approx0.13$ is the $x$-axis ($\phi=0$).
Above the critical energy this basin boundary bifurcates into three
`branches', the upper (with $\phi>0$) corresponding to the
unstable periodic orbit $\Pi_3$, the middle corresponding to
$\Pi_0$ and the lower corresponding to $\Pi_2$. Moreover, the basin
boundaries above the critical energy $E_0$ are fractal. For example, the
function $j(\phi)$ shown in the previous Fig.~3 (which is clearly
fractal) can be obtained as a section of Fig.~4 for $E=0.16$.
Fig.~4 is also suitable for a better localization of the critical energy
$E_0$ (from Fig.~3 it was only clear that $E_0<0.16$). More
detailed numerical calculations (not presented here) led us to the
value $E_0=0.1364\pm 0.0001$.

The chaotic part of the medium energy region ($E_0<E<{1\over 6}$)
in Fig.~4 contains two interesting subregions which deserve
special attention: energies just below $E={1\over 6}$ and
energies just above $E=E_0$. These subregions will be discussed
below.

Fig.~5 shows an enlargement of Fig.~4 for energies close to
$E={1\over 6}$. It is obvious that there is another bifurcation
at $E_1\approx 0.165$ which is somewhat similar to that which occured at
$E=E_0$. The fractal structure above $E_1$ becomes even more complex,
roughly speaking with five peaks in $\tau_s(\phi)$ at each fractal
level, instead of three peaks shown in Fig.~3.

However, Fig.~5 indicates more:
as the energy $E$ approaches ${1\over 6}$, at least one other bifurcation
occurs. This naturally leads us to the following {\it conjecture}:
There is a infinite number of bifurcations in the fractal basin
boundary below $E={1\over 6}$ at each of which the basin boundary
along $\phi=0$
bifurcates into three branches.

According to the conjecture, the density of bifurcation points
grows to infinity as $E\to{1\over 6}$, where the third exit (to
$x\to-\infty$) is opened. The periodic orbit $\Pi_0$ vanishes while
a periodic orbit $\Pi_1$ in the new `neck' of the potential is created. Above
$E={1\over 6}$ there are three exits to infinity and three
corresponding unstable periodic orbits $\Pi_1$, $\Pi_2$, $\Pi_3$,
similar to the structure for the monkey-saddle system\footnote{There are
two other
periodic orbits $\Pi_4$ and $\Pi_5$ in the monkey-saddle system,
which
are more difficult to localize. We do not discuss
their analogues in our system.} (see \cite{Rod},
\cite{PoVe2}).

It would be interesting to investigate the transition region around
the first bifurcation point $E=E_0$ in detail. However, it proved to be very
demanding on the accuracy of numerical methods. Fig.~6 indicates
why: typical orbits in this energy region reach infinity at
$\tau_s\gg 1$, after `bouncing' many times between the potential walls
($\tau_s\approx 400$ in Fig.~6, which is still moderate: we were
able to follow number of orbits with $\tau_s>10^4$). The orbit shown in
Fig.~6 started at $\phi=0.002$ with $E=0.13669$ and performed 110
bounces before escaping to $j=2$. The numerical computation is  extremely
sensitive to the accuracy of the method. Such sensitivity is
present close to any fractal basin boundary shown in Fig.~4, but
it is usually restricted to a tiny range of initial conditions.
However, it is present in the whole subregion bordered roughly by
the rectangle $E_0<E<E_0+0.0005$, $|\phi|<0.003$. This region near the first
bifurcation and similar (but smaller) regions near the higher
bifurcations represent orbits with the maximum level of
unpredictability in the system under consideration.

\section{The box-counting dimension}

Finally, we present a more complete characterization of the topology
of chaotic orbits in the medium energy region.
In  discussion so far, we have restricted our analysis to
specific (yet important) orbits starting from {\it rest from}
$B_r$. A completely general orbit can start at an arbitrary
(not forbidden) point with given energy $E$ and direction of the
initial impulse. For our purposes it is crucial to investigate
the {\it character} of the basin boundaries, i.e. their regular
vs. fractal nature. All unbounded orbits starting at points $(x,y)$
with $y>0$ (or $y<0$) {\it which never cross} the manifold
section $y=0$ has the same certain exit $j=3$ (or $j=2$,
respectively). Therefore, the basin {\it boundaries} can appear
only among those orbits which cross $y=0$. In order to obtain
the basin boundary portrait it is thus sufficient to systematically
study orbits which start from the (negative) $x$-axis,
i.e. from points $P=(x\le0, y=0)$, where $x\in[x_{min}(E), 0]$,
with initial impulse $p_x, p_y$.  (A complementary discussion of orbits
starting at $x\ge0, y=0$ would be analogous.) These orbits belong to
the energy manifold $H={1\over 2}(p_x^2+p_y^2)+V_m(x,0)=E$. The set
of initial conditions considered can thus be represented by a disc
in the $(p_x,p_y)$ plane in which each point characterize a specific
initial velocity $\dot x=p_x, \dot y=p_y$ of an orbit starting at
$(x,0)$, where $x\le0$ is the solution of the equation $x^2+{2\over
3}x^3=2E-p_x^2-p_y^2$. The circular boundary of the disc
corresponds to geodesics starting at $x=0$ with maximum permitted
impulse $p_x^2+p_y^2=2E$, whereas the center of the disc denotes
$p_x=0=p_y$ and $x_{min}(E)\in(-1,0)$. In Fig.~7 we present typical
discs for $E=0.16$ (left) and $E=0.166$ (right) in which each point
representing specific initial data has again been colored in black or
white according to the exit (`up' and `down', respectively)
reached by the corresponding evolved orbit. Analogous discs in
the monkey saddle potential can be found in \cite{Rod},
\cite{PoVe2}.

It is obvious that the discs contain fractal basin boundaries,
patterns and bifurcations which resemble those shown in Fig.~4
and Fig.~5. Interestingly, the geometrical pictures are very similar
despite the fact that they describe different situations. In
Figs.~4 and 5 the main independent parameter is $E$, whereas $E$
is fixed in Fig.~7 with the main parameter being $x$.

The fractal structure of the basin boundaries can rigorously be
characterized by the box-counting dimension \cite{BGOB}, \cite{Ott}.
A large number of randomly chosen initial conditions represented by
points in the above discs is evolved. For each point $P$ its
neighbourhood of radius $\varepsilon$ is also inspected. If not
all the points in the neighbourhood belong to the same basin as
$P$ then the point is denoted as `uncertain'. The fraction
$P(\varepsilon)$ of uncertain initial conditions among all chosen
points (we simulated 10000 points for each $\varepsilon$) scales
as $P(\varepsilon)\sim\varepsilon^{2-d}$, where the parameter $d$
gives the box-counting dimension of the fractal basin boundaries
of the discs shown in Fig.~7. Numerical simulation indicates
that $d=1.34\pm0.07$ for $E=0.16$, and $d=1.43\pm0.06$ for
$E=0.166$. In fact, the dimension grows with $E$ from $d=1$ for
$E<E_0\approx0.1364$ to $d\to1.50\pm0.05$ as $E\to{1\over6}$.

\section{Concluding remarks}

We have discussed in detail the Hamiltonian system with the
potential (\ref{E6}) which is a simple but interesting modification
of the famous H\'enon-Heiles potential (\ref{VHH}). Interestingly, orbits in
the system studied correspond to the geodesic motion of free test particles
in the space-times describing plane-fronted gravitational waves
with both homogeneous and non-homogeneous components in the
structural function (\ref{f23}). We have demonstrated that for
energies $E>E_c\approx0.1364$ the motion becomes chaotic since
the basin boundaries separating two possible exits to infinity
have a fractal structure. For $E>{1\over6}$ the third exit opens
and the resulting chaotic behaviour has a similar structure as
in the monkey saddle potential $V_{ms}$ described in \cite{Rod},
\cite{PoVe2}. This `pure cubic' potential for the monkey
saddle system is obtained by setting $\alpha=0$ in (\ref{f23}).
On the other hand, if the cubic term in the function $f$ vanishes
($\beta=0$), the gravitational wave is homogeneous and the
equations of geodesic motion become linear, i.e. integrable.

\section*{Acknowledgments}

We acknowledge the support of grants GACR-202/99/0261 from the Czech
Republic and GAUK~141/2000 of the Charles University. We thank the
developers of the software system FAMULUS which we used for computation
and drawing of all the pictures. Also, we thank Jerry Griffiths
for his help with the manuscript.

\newpage

\section*{Figure Captions}

\begin{description}

\item{Fig.~1\ } The equipotentials $V_m(x,y)=\textstyle{\frac{1}{2}}
(x^2-y^2)+\textstyle{\frac{1}{3}} (x^3-3xy^2)=E$\  for nine values of
energy $E$ representing all typical situations. The forbidden regions
are grey.

\item{Fig.~2\ } An example of orbits starting from rest from the `right
branch' of the level curve $B_r$ with energy $E=0.16$. The initial
positions on $B_r$ are parametrized by an angle $\phi$. The orbits
$\Pi_0$ and $\Pi_3$ are unstable periodic orbits.

\item{Fig.~3\ } Plots of the function $j(\phi)$ labelling
the two possible exits to infinity and the function $\tau_s(\phi)$
denoting the time at which the orbit starting at given $\phi$ reaches
infinity ($E=0.16$ here). The enlargements up to the fourth level
confirm the fractal structure of basin boundaries.

\item{Fig.~4\ } The basins of the two exits, `up' (black) and `down'
(white), in the subset of initial conditions corresponding to orbits
starting from $B_r$. These are uniquely characterized by energy $E$
and starting angle $\phi$.

\item{Fig.~5\ } Enlargement of Fig.~4 for energies close to
$E=\textstyle{\frac{1}{6}}$. The basin boundaries are fractal.

\item{Fig.~6\ } Typical geodesics ($\phi=0.002$, $E=0.13669$)
`bounce' many times between the walls before escaping to
infinity.

\item{Fig.~7\ } The discs $(p_x, p_y)$ for $E=0.16$ and $E=0.166$
show the topology of orbits in the corresponding energy manifold.
Each point in the disc represents an orbit starting at specific
point $x<0$, $y=0$ with a given velocity. Its colour denotes one
of the two possible exits. Again, the fractal basin boundaries
indicate chaos.

\end{description}

\end{document}